\documentclass[twocolumn,superscriptaddress,secnumarabic,
amssymb,nobibnotes,aps,prd,nofootinbib]{revtex4}
\usepackage{graphicx}
\usepackage{epsf}
\usepackage{bm}
\usepackage{amsfonts}
\usepackage{amssymb}
\usepackage{epstopdf}
\usepackage{natbib}%
\usepackage{color}
\usepackage{amsthm,amssymb,color}
\usepackage{amsfonts,latexsym, hyperref, bbm, mathrsfs}
\begin{document}
	
\title{Cosmological  Time Crystal: \\ Cyclic Universe with a  small $\Lambda$ in a toy model approach}
	
\author{Praloy Das}
\email{praloydasdurgapur@gmail.com}
\affiliation{Physics and Applied Mathematics Unit, Indian Statistical
	Institute, 203 Barrackpore Trunk Road, Kolkata-700108, India}

\author{Supriya Pan}
\email{span@research.jdvu.ac.in}
\affiliation{Department of Mathematics, Raiganj Surendranath Mahavidyalaya, Sudarshanpur, Raiganj, Uttar Dinajpur, West Bengal 733134, India}

\author{Subir Ghosh}
\email{subirghosh20@gmail.com}
\affiliation{Physics and Applied Mathematics Unit, Indian Statistical
	Institute, 203 Barrackpore Trunk Road, Kolkata-700108, India}

\author{Probir Pal}
\email{probir.kumarpal@gmail.com}
\affiliation{Department of Physics, Barasat Government College,
	10, KNC Road, Barasat, Kolkata 700124, India}

\begin{abstract} 
A new form  Time Crystal has been proposed and some of its consequences have been studied. The model is a generalization of the Friedmann-Robertson-Walker (FRW) cosmology endowed with noncommutative geometry corrections. In the mini-superspace approach the scale factor undergoes the time periodic behavior, or Sisyphus dynamics, which allows us to interpret this Cosmological Time Crystal as a physically motivated toy model to simulate cyclic universe. 
Analyzing  our model purely from 
Time Crystal  perspective reveals many novelties such as a  complex singularity structure  (more complicated than the previously encountered swallowtail catastrophe) and a richer form of Sisyphus dynamics.
In the context of cosmology, the system can serve as a toy model in which, apart from inducing a form of cyclic universe feature,  it is possible to generate an arbitrarily small positive effective Cosmological Constant. We stress that the model is purely geometrical without introduction of  matter degrees of freedom.

\end{abstract}

%----------------------------------------
\maketitle
%---------------------------------------------------	
 \section{Introduction}

In this paper we aim to  apply the fascinating concept of Classical  Time Crystal (CTC), proposed by Shapere and Wilczek \cite{wil,wil11} (see \cite{rev} for a recent review), in an extended model of Friedmann-Robertson-Walker (FRW) cosmology. Specifically, the extension is induced by Non-Commutative (NC) gravity contribution with an   underlying Quantum Gravity perspective. It was derived by Fabi, Harmes and Stern \cite{stern}. In a nutshell two of our principal results are the following:\\
 (i) The scale factor borrows the Sisyphus like periodic behavior that characterizes the CTC but, more importantly for our present interest, it can  naturally serve  as a physically motivated toy model for a  cyclic universe, conceived  by Steinhardt and Turok  \cite{cycl}~{\footnote{We note that the perspectives of cyclic cosmology \cite{cycl} and that in the present model is somewhat different though since in the former, Quantum Gravity effects are not considered to be significant whereas in the latter the cyclic cosmological features emerge due to the noncommutative contributions which in turn are generally thought to be induced by  Quantum Gravity effects. We consider a closed universe.}}
 \\(ii) Once again borrowing a CTC feature, the minimum energy state (or ground state) consists of a condensate leading to an arbitrarily small positive cosmological constant $\Lambda$. 
 
 Furthermore, it needs to be stressed that our model is purely geometric in the sense that no matter degrees of freedom are added from outside. This should be contrasted with recent works in cosmological CTC \cite{Bains} where a scalar field model with eternal oscillations in an expanding FRW spacetime was discussed (see also \cite{eass} for further developments on the model). From TC perspective as well there is some novelty since recent works \cite{th} concerning physical realization of TC are all in the quantum domain {\footnote{The quantum TC was proposed by Wilczek \cite{wil2} with experimental models for quantum TC in \cite{ex}.}} whereas our framework is purely classical. The only classical example studied so far is in \cite{wil,wil11} that is not very realistic.

It is worthwhile to mention a remarkable coincidence concerning the present work with two earlier observations: from the time-symmetric nature of the NC metric, in \cite{stern} itself the  authors had suggested that the NC FRW metric might describe a bouncing universe from a restricted study of geodesics. This is not connected to the idea of TC since the latter appeared much later. On the other hand, in a recent article on TC \cite{zak} Zakrzewski has conjectured that cyclic evolution of cosmology may emerge as a form of TC resulting from spontaneous breaking of time translation invariance but did not mention any specific scenario. Our work provides a successful union of these two ideas in \cite{stern} and \cite{zak}.

 We have introduced three concepts, Time Crystal, Cyclic Universe and non-commutative geometry corrected General Relativity or gravity. Let us elaborate on these briefly.
 
 {\it{Classical Time Crystal}}: Within a few years of the theoretical conjecture, TC  has created an enormous amount of interest, both in theoretical \cite{th} and experimental \cite{ex} contexts (see \cite{rev} for a recent review). 
 
 Symmetry of a system is spontaneously  broken when  the ground state (or classically, the lowest energy state) of a system is less symmetrical than the equations of motion that control  the system. Prominent examples from the quantum world are  the Higgs boson, ferromagnets and antiferromagnets, liquid crystals, and superconductors. Examples in classical systems are rare.

 The name Time Crystal is borrowed from the familiar (space) crystal that has a spatially ordered structure in its ground state. It is a manifestation of breaking of continuous translation symmetry, leaving behind a ground state  with discrete translation symmetry~{\footnote{It was shown by one of us in \cite{sg} that spatial higher derivative terms in a field theory can induce breaking of spatial translation invariance (as an alternative to Shapere-Wilczek scheme) and it can be extended to break time translation invariance as well.}}. The question posed by  Shapere and  Wilczek \cite{wil,wil11} is the following: can a system with {\it{time}}-periodic ground states that breaks continuous {\it{time}} translational  invariance exist? The resulting system with discrete time translation symmetry was referred to as Time Crystal. A mathematical model for a classical TC was provided in \cite{wil} and later in \cite{wil11} a more ``physical'' model was constructed in which the relevant degree of freedom undergoes periodic Sisyphus dynamics in its lowest energy state. The latter model in a certain limit reduces to the strictly CTC model. Indeed, the strict CTC inherently has singular behavior whereas the physical realization is throughout well behaved. These will be discussed in more detail later but essentially CTC has a non-canonical form of kinetic energy that minimizes (not for zero velocity as in conventional systems but) at a non-zero velocity, much in analogy to Spontaneous Symmetry Breaking (SSB) where a potential function  minimizes at a non-zero (but constant)  value of the field variable that constitutes the ground state condensate. A subtle but essential point is that for time crystal behavior the condensates must show spatial or temporal non-uniform (but periodic) behavior.  
 
 Even though it will not concern us here, the quantum TC proposed by Wilczek \cite{wil2} had certain drawbacks as pointed out in \cite{bruno} and improved versions of the quantum TC appeared later \cite{th} and TC characteristics  have been observed in laboratory experiments \cite{ex} (for a recent review see \cite{rev}).
 
 It is reassuring to emphasize that TCs do not break any of the sacred thermodynamic principles such as for example the second law of thermodynamics even though TCs may appear to simulate a perpetual motion machine. The key point is to note that the movement in ground state does not involve any external work and no usable energy can be extracted from the motion in ground state. In fact, TCs can have interesting applications such as   precise timekeeping,   simulation of ground states in quantum computing schemes \cite{zak}, and finally,  in cyclic  cosmology, as proposed in the present work.

 {\it{Cyclic Universe}}: Cosmological models with a classical non-singular bounce was proposed in \cite{stein, cycl} as an alternative to inflation \cite{inf}. A few important and generic positive features of the bouncing cosmology model are the following \cite{stein}: it avoids the cosmic singularity problem;  resolve the horizon problem;
 explain the smoothness and flatness of the universe as well as  the small entropy at the starting of  expanding phase of the universe. The Big Bang  singularity is substituted by a bounce wherein  the scale factor contracts to  a   finite  size that is  well above the Planck length, and rebounds to the expanding phase. This provides the geodesic completeness (that is absent in the Big Bang scenario) since every co-moving particle  has a worldline that can be traced backwards through the bounce arbitrarily far back in time. This makes the energy density below the Planck density at all  times so that   quantum gravity effects are expected not to play a major role. Hence, the need to explain  quantum-to-classical transition does not arise. Causal connectedness, a generic feature of bouncing scenario, is  achieved in the contracting phase backwards in time by passing through the bounce. The flatness of the universe is explained by noting that  the cosmic curvature factor is exponentially suppressed at  the time the universe reaches the bounce, generating an extremely small  spatial curvature that agrees with the flatness observed at present. Lastly,  bouncing models satisfy the second law of thermodynamics. It has a small entropy density, tiny in comparison to the maximum entropy possible, meaning that it was still less  at the beginning of the expansion phase. This is possible because the present  observable universe  was just an infinitesimal fraction of the horizon size long before the bounce possessing a small amount of entropy available.
 
  The cyclic model  has turned out to be a popular and well studied model of our universe \cite{Novello} where the universe passes through an infinite sequence of expanding and contracting phases without having a beginning or ending of time. This is also in agreement with the astronomical data \cite{Lehners}. 
  
 However, the challenge remains in establishing the {\it{mechanism}} that accounts for
 the bounce explicitly in a cosmological model.  It is here that our cosmological time crystal model (or, NC corrected FRW model)
 can throw some light since the dynamics of  periodic expansion and contraction ($\sim$ Sisyphus dynamics) of the scale factor is demonstrated explicitly in our work. 

 {\it{General Relativity in noncommutative spacetime}}: 
  NC geometry or equivalently a generalized form of canonical (Heisenberg) phase space algebra was introduced long ago by Snyder \cite{sn} to ameliorate short distance divergence but that idea was not successful.  The seminal work of Seiberg and Witten \cite{sw} (see review works \cite{ncrev}) put NC geometry in the limelight by showing that string theory in certain low energy limits can be identified with quantum field theory extended to NC spacetime of the form
\begin{equation} 
[x_\mu,x_\nu]=i\Theta_{\mu\nu}
\label{fst} 
\end{equation}
with constant antisymmetric NC parameter $\Theta_{\mu\nu}=-\Theta_{\nu\mu}$. However, due to a subtle effect, known technically as ultraviolet-infrared mixing, the short distance scale $\Delta x_\mu\Delta x_\nu\approx {\sqrt{\theta_{\mu\nu}}}$ induced from (\ref{fst}) does not improve the ultraviolet behavior of quantum field theory. The systematic procedure of extending a field theoretic model living in conventional spacetime to NC spacetime is to replace local products of fields by
the Groenewold-Moyal star product   given by
\begin{equation} \star = \exp\;\biggl\{ \frac {i}2 \Theta^{\mu\nu}\overleftarrow{
	\partial_\mu}\;\overrightarrow{ \partial_\nu} \biggr\} \;,\label{k6} 
\end{equation}  
where $\overleftarrow{\partial_\mu}$ and $\overrightarrow{ \partial_\mu}$ are left and right derivatives, respectively, with respect to  some generic coordinate $x^\mu$. To be explicit, products of local fields like $A(x)B(x)$ are replaced by 
$$A(x)\star B(x)=A(x) \exp\;\biggl\{ \frac {i}2 \Theta^{\mu\nu}\overleftarrow{
	\partial_\mu}\;\overrightarrow{ \partial_\nu} \biggr\}  B(x)$$
in the action and construct NC extended equations of motion. The NC contributions will appear as a power series in $\theta_{\mu\nu}$ obtained from expanding the $\star$-product formally as a power series in  $\theta_{\mu\nu}$. Furthermore, conventional gauge theories require a special treatment: the Seiberg-Witten map connects NC degrees of freedom to conventional  degrees of freedom \cite{sw}. Let us now discuss the motivation for NC gravity. 

At present, there is no  universally accepted  theory (backed by  experiments), that can describe physics at or below Planck length. It is argued that in this regime quantum effects cannot be ignored.  Various attempts to formulate such a theory is generically referred to as examples of Quantum Gravity theories. 
It has been established  
that classical General Relativity in conjunction with Quantum Mechanics demands 
existence of a fundamental length scale \cite{hsu}. Examples of a class of such theories are NC field theories mentioned earlier. Thus, it is only natural that NC extension of General Relativity is sure to serve as a strongly motivated toy model to simulate  Quantum Gravity. However, as noted in the text, further and more elaborate analysis is needed before the NC gravity can pose as a  candidate for  Quantum Gravity.

The above discussion puts our paper in its proper perspective.  We have  studied a physically motivated toy model, NC FRW \cite{stern}, in the light of   CTC, in  a closed universe.  The results are indeed striking $-$ in the 
 Mini-Superspace toy model approach, we find that the NC effects induce a SSB in time translation invariance that results in a cyclic type of cosmology. Furthermore, we provide a natural way of generating a {\it{small}} positive cosmological constant. We emphasize that, contrary to the existing cyclic models that require a potential term, our model is geometric and does not require any matter from outside.  To the best of our knowledge, our work provides the first instance of a realistic and physically motivated Classical Time Crystal model that is a generalization  of the simplest forms of  CTC model of \cite{wil, wil11}.

 The paper is organized as follows: In Section \ref{sec-2}, a brief and self contained discussion on generic classical time crystal is provided. At the same time the model studied has direct relevance with our work concerning the cosmological time crystal, the latter being richer in structure. In Section \ref{sec-3}, the noncommutative gravity is introduced and the noncommutative metric structure is derived. Section \ref{sec-4} deals with the noncommutative mini-superspace action. This constitutes the major part where the explicit cosmological time crystal models are derived. In Section \ref{sec-5}, the cosmological time crystal is analyzed in detail.  Section \ref{sec-6} discusses the interesting result that the cosmological time crystal can yield an effective small positive Cosmological Constant. In Section \ref{sec-6}, a noncommutative generalization of Friedman equation is derived. We conclude with a discussion and future directions of work in Section \ref{sec-8}.

\section{Generic model of Classical Time Crystal (CTC)}
\label{sec-2}
Let us start by observing a generic CTC in more detail. From Hamiltonian dynamics point of view it seems that CTC cannot exist. Consider a Hamiltonian $H(p,\phi)$ with coordinate $\phi$ and conjugate momentum $p$. The Hamiltonian $H(p,\phi)$ minimizes at $\frac{\partial H}{\partial p}=\frac{\partial H}{\partial \phi}=0$. On the other hand,   the Hamilton's equations of motion states that  $\dot\phi = \frac{\partial H}{\partial p}$. Putting together we get for the minimum energy state $\dot\phi = \frac{\partial H}{\partial p}=0$ indicating that $\phi $ should be a constant. Thus, classical ground state should be static contrary to a CTC ground state.

However, the  problem is more subtle since in CTC the canonical momentum $p$ leads to a multivalued Hamiltonian as a function of $\dot \phi$ with cusps at $\frac{\partial p}{\partial \dot \phi}=0$, where the Hamiltonian equations of motion are not valid. Thus, it might be possible for CTC to avoid the negative conclusion mentioned above. We discuss an explicit example \cite{wil,wil11} that will show how  the system ground state  can adjust itself to the contrasting demands of a time invariant (constant position) and simultaneously time varying (constant velocity) state.

The idea is analogous to the phenomenon of SSB, albeit in velocity (or momentum) space. In  conventional SSB the energy minimizes at zero velocity but   the potential energy $V(\phi )$ has  its minimum   at a non-zero but constant 
$\phi=\phi_0$.	However, in \cite{wil,wil11} the authors introduced a simple dynamical model where the minimum value of  kinetic energy is obtained at a time dependent $\phi(t)$ (but constant $\dot\phi(t)=\dot\phi_0$) so that even classically the ground state condensate has a motion. In conventional SSB, for the Lagrangian $\mathcal{L}=\frac{1}{2} \dot {\phi}^2 -V(\phi )$, if the potential $V(\phi)=-\frac{A}{2}\phi^2+\frac{B}{4}\phi^4$ has a minimum at $\phi_0=\pm {\sqrt{\frac{A}{B}}}$,  the energy $E=\frac{1}{2} \dot {\phi}^2 +V(\phi )$ also minimizes at constant $\phi_0$.

Let us now return to CTC. Consider  a generic Lagrangian\footnote{In our cosmological TC model the Lagrangian is more involved but has similar features.}
\begin{equation}
 \mathcal{L}=-\frac{A}{2}\dot{\phi}^2+\frac{B}{4}\dot{\phi}^4 -V(\phi )
\label{01}
\end{equation}
that leads to the energy
\begin{equation}
E=-\frac{A}{2}\dot{\phi}^2+\frac{3}{4}B\dot{\phi}^4+V(\phi ).
\label{02}
\end{equation}
Rewriting the energy as 
\begin{equation}
E=\frac{3B}{4}(\dot \phi^2-\frac{A}{3B})^2-\frac{A^2}{12B} +V(\phi),
\label{001}
\end{equation}
it is clear that  this energy  minimizes at 
\begin{equation}
\dot \phi_0 =\pm \sqrt{\frac{A}{3B}},~~\phi=\phi_0.
\label{03}
\end{equation}
This is CTC ground state with mutually opposing requirements of simultaneous constant (non-zero) velocity and constant position (For an alternative approach see \cite{sg}). 

To further elucidate the situation we note that the Lagrangian equation of motion
\begin{equation}
\ddot \phi = -\frac{\frac{\partial V}{\partial \phi}}{3B(\dot \phi ^2-\frac{A}{3B})},
\label{04}
\end{equation}
diverges at the energy minima. Furthermore, energy becomes a multivalued 
function of $p$,
\begin{equation}
p=\frac{\partial \mathcal{L}}{\partial\dot{\phi}}=-A\dot \phi +B\dot \phi ^3,
\label{05}
\end{equation}
with cusps at the energy minima,
\begin{equation}
\frac{\partial p}{\partial \dot\phi}=-A+3B \dot \phi^2 =0 ~\rightarrow ~ \dot \phi_0 =\pm \sqrt{\frac{A}{3B}}. 
\label{06}
\end{equation}
Later we will show explicitly that the cusp structure leads to catastrophe for our cosmological TC.

Hence, at exactly ground state the system becomes singular but interesting behavior is recovered for the system being arbitrarily close to the ground state with energy $E$ slightly above the minimum energy, $E=E_{min} +\Delta = -\frac{A^2}{12}+\Delta$, $\Delta$ being small. Assuming this energy occurs at $\phi_t$, one can approximately solve for $\dot \phi$ \cite{wil} to obtain 
\begin{equation}
\dot \phi =\pm {\sqrt{\frac{A}{3B}}} \pm {\sqrt {\frac{1}{A}(\frac{\partial V}{\partial \phi})_{\phi_t}(\phi_t-\phi)}}.
\label{07}
\end{equation}
The above equation (\ref{07}) has four independent solutions. To interpret the behavior of the system close to the ground state we note that $\dot \phi$ tends to remain close to either of $ \pm {\sqrt{\frac{A}{3B}}}$ but it gets altered by the second term. The latter, however, cannot go beyond a certain value since $\phi$ can not exceed $\phi_t$. Altogether the position changes (due to one of the non-zero velocities $ \pm {\sqrt{\frac{A}{3B}}}$), but when it reaches near about the value that minimizes the potential, the velocity abruptly falls back and this cycle is  repeated $-$ in short the Sisyphus dynamics. Thus the system is  force  to undergo ``{\it Sisyphus dynamics}'' \cite{wil11} where on an average $\phi $ stays close to the constant $\phi_0$ but periodically goes through phase of non-zero  $\dot \phi =\dot\phi_0$ (constant).

In order to visualize the Sisyphus dynamics of the ground state in \cite{wil11}, the authors have provided an alternative mechanism. Physically this model can be interpreted as a charge moving in the plane in presence of an electric field. A new parameter with an auxiliary variable is introduced that results in a well behaved system. The singularity reappears in the vanishing limit of the new parameter. In principle this parameter can be arbitrarily small and numerically it does not alter the system properties significantly. One can see the Sisyphus dynamics in the ground state motion in this enlarged (regularized) set up. We will discuss this scheme for our cosmological TC.

\section{Gravity in Non-Commutative space and Non-Commutative FRW}
\label{sec-3}

Although NC non-abelian gauge theories have been successfully constructed using Seiberg-Witten map, NC generalization of General Relativity poses problems mainly because  it is difficult to impose invariances such as general coordinate covariance and local
Lorentz symmetry in NC spaces. Furthermore, torsion-free  derivatives that satisfy the metricity condition  are difficult to define on NC spaces.

Out of several inequivalent approaches we will utilize the proposal of Fabi et. al. \cite{stern} who exploit an earlier work \cite{cal} that follows the approach in  \cite {chams} where Seiberg-Witten map is utilized to construct NC analogue of $SO(4,1)$ gauge theory and subsequently impose Wigner-Inonu contraction to make contact with General Relativity. A further simplification is considered \cite{ncgr}: instead of deriving the NC dynamical equations for the metric and solving them one uses the existing solutions of Einstein equation in normal spacetime and appropriately extends them to NC spacetime in the above framework. Possible limitations of this scheme is mentioned in \cite{stern}. The NC FRW construction has been achieved in this way in \cite{stern} which we now briefly outline.

General Relativity, as a gauge theory, is expressed in terms of  spin connection and vierbein  one-forms,  $\omega^{ab}=-\omega^{ba}$ and $e^a$,  respectively. Here, the  Lorentz indices are   $a,b,...=0,1,2,3$ that act in the flat metric space $\eta={\rm diag} (-1,1,1,1)$. On the other hand,  the space-time metric is given by
\begin{equation} g_{\mu\nu}= e^a_{\;\;\;\mu}e^b_{\;\;\;\nu}\eta_{ab} \;.\label{k1} \end{equation}
Infinitesimal local $ISO(3,1)$ transformations of  $\omega^{ab}$  and $e^a$ appear as 
\begin{eqnarray}
\delta \omega^{ab} &=& d\lambda ^{ab} +[\omega,\lambda]^{ab} \cr  & &\cr
\delta e^{a} &=& d\rho^{a} +  \omega^a_{\;\;\;c}\rho^{c} - \lambda ^a_{\;\;\;c} e^{c} \;,\label{k2}
\end{eqnarray}
for infinitesimal parameters   $\lambda ^{ab}=-\lambda ^{ba}$ and $\rho^a$, and where $[\omega,\lambda]^{ab}= \omega^a_{\;\;\;c}\lambda ^{cb} - \lambda^a_{\;\;\;c}\omega ^{cb}$.  The spin curvature and torsion two-forms, $R^{ab}=-R^{ba}$ and $T^a$ , respectively, are obtained  from  $\omega^{ab}$ and $e^a$  in the conventional way:
\begin{eqnarray}  R^{ab}&=&d\omega^{ab} +
\omega^a_{\;\;\;c}\wedge\omega^{cb} \cr & &\cr
T^a&=&de^a
+\omega^a_{\;\;\;b} \wedge e^b \; .
\label{k3}
\end{eqnarray}
The complete details are provided in \cite{stern}. The above $ISO(3,1)$ gauge theory is the Wigner-Inonu contraction of $SO(4,1)$ gauge theory consisting of   potential one-form $A^{AB}$ and  curvature two form $F^{AB}=-F^{BA}$  are
\begin{equation} F^{AB}=dA^{AB} +
A^A_{\;\;\;C}\wedge A^{CB}\;. 
\label{k4}
\end{equation}
where $A,B,..=0,1,2,3,4$ living in a space with metric tensor $diag (-1,1,1,1,1)$.  The contraction is performed  by imposing 
\begin{eqnarray} \Lambda^{ab} =\lambda ^{ab} &\qquad & \Lambda^{a4} = \kappa \rho^a \cr & &\cr
A^{ab} =\omega^{ab} &\qquad & A^{a4} = \kappa e^a \cr & &\cr
F^{ab} =R^{ab} &\qquad & F^{a4} = \kappa T^a \;
\label{k5}
\end{eqnarray}
in the limit $\kappa\rightarrow 0$. 

NC  version of the  $SO(4,1)$ curvature two form is 
\begin{equation} \hat F^{AB}=d\hat A^{AB} +
\hat A^A_{\;\;\;C}\stackrel{*}{\wedge} \hat A^{CB}\;. 
\label{k7}
\end{equation}
$\stackrel{*}{\wedge}$ denotes an exterior product   where
the local product terms are  replaced by $\star$-product.
The NC analogues of spin connection,  vierbein, curvature and torsion forms, denoted respectively  by  $\hat \omega^{ab}$, $\hat e^a$, $\hat R^{ab}$ and $\hat T^a$ follow from  $\hat A^{AB}$ as before,
\begin{eqnarray} 
\hat A^{ab} =\hat \omega^{ab} &\qquad & \hat A^{a4} = \kappa \hat e^a \cr & &\cr
\hat F^{ab} =\hat R^{ab} &\qquad & \hat F^{a4} = \kappa \hat T^a \;, \qquad {\rm as}\; \kappa\rightarrow 0\;.\label{k8}
\end{eqnarray}	

Seiberg-Witten map comes into play when  $\hat A^{AB}(x,\Theta)$, $\hat F^{AB}(x,\Theta)$ and $\hat \Lambda^{AB}(x,\Theta)$ are expanded as a series in powers of  $\Theta^{\mu\nu}$
\begin{eqnarray} 
\hat A^{AB}_\mu (x,\Theta) &=& A^{AB}_\mu(x)\;+\;\left.\matrix{ \cr A^{AB}_{\mu}\cr{}^{ (1)}\cr}\right.(x,\Theta)\;+\;\left.\matrix{ \cr A^{AB}_{\mu}\cr{}^{ (2)}\cr}\right.(x,\Theta)\;+\;\cdots\cr & &\cr
\hat F^{AB}_{\mu\nu}(x,\Theta) &=& F^{AB}_{\mu\nu}(x)\;+\;\left.\matrix{ \cr F^{AB}_{\mu\nu}\cr{}^{ (1)}\cr}\right.(x,\Theta)\;+\;\left.\matrix{ \cr F^{AB}_{\mu\nu}\cr{}^{ (2)}\cr}\right.(x,\Theta)\;+\;\cdots\cr & &\cr
\hat \Lambda^{AB}(x,\Theta) &=& \Lambda^{AB}(x)\;+\;\left.\matrix{ \cr \Lambda^{AB}\cr{}^{ (1)}\cr}\right.(x,\Theta)\;+\;\left.\matrix{ \cr \Lambda^{AB}\cr{}^{ (2)}\cr}\right.(x,\Theta)\;+\;\cdots\;
\label{k9}
\end{eqnarray}
The numerics $(1),(2),..$ in the right hand side denotes the order of $\Theta$. In a similar fashion NC generalizations of vierbeins and spin connections are expressed as  power series expansion in $\theta$:
\begin{eqnarray}
\hat e^{a}_\mu (x,\Theta) &=& e^{a}_\mu(x)\;+\;\left.\matrix{ \cr e^{a}_{\mu}\cr{}^{ (1)}\cr}\right.(x,\Theta)\;+\;\left.\matrix{ \cr e^{a}_{\mu}\cr{}^{ (2)}\cr}\right.(x,\Theta)\;+\;\cdots
\cr & &\cr \hat \omega^{ab}_\mu (x,\Theta) &=& \omega^{ab}_\mu(x)\;+\;
\left.\matrix{ \cr \omega^{ab}_{\mu}\cr{}^{ (1)}\cr}\right.(x,\Theta)\;+\;\left.\matrix{ \cr \omega^{ab}_{\mu}\cr{}^{ (2)}\cr}\right.(x,\Theta)\;+\;\cdots\;.\label{k10}
\end{eqnarray}	

{\it{Noncommutative FRW}}: As stated earlier, the above scheme allows us to compute NC extensions of known metrics in a consistent way that respects the gauge invariances of General Relativity since the power series constructions are obtained from the Seiberg-Witten map. This finally brings us to the NC FRW metric.

The  conventional Friedmann-Robertson-Walker metric reads,
\begin{equation} ds^2= -dt^2 + a(t)^2\Bigl(\frac{dr^2}{1-kr^2} +
r^2(d\theta^2 + \sin^2 \theta \;d\phi^2)\Bigr)
\;, \label{k11}
\end{equation} where $a(t)$ is the scale factor. Vierbein one forms are identified as,
\begin{equation}
e^0 = dt\;, e^1 = \frac {a(t)\; dr}{\sqrt{1-kr^2}}\;,  e^2 = a(t) r\; d\theta\;,  e^3 =a(t) r \sin\theta \;d\phi \;. \label{k12}
\end{equation}
In order to make torsion zero, the  following choice of  spin connection one forms are required,
\begin{widetext}
\begin{equation}
\left.\matrix{& & \cr \omega^{01} = \chi dr\;, & \omega^{02} = \dot a r \;d\theta\;, & \omega^{03}= \dot a r \sin\theta \;d\phi\cr & & \cr \omega^{12} = -\sqrt{1-kr^2} \;d\theta &  \omega^{31} =\sqrt{1-kr^2} \sin\theta \;d\phi &  \omega^{23} =-\cos\theta\; d\phi \cr& & \cr}\right.\label{k13}\;,
\end{equation}
\end{widetext}
where the dot denotes time derivative.  Constructing the   curvature scalar 
${\cal R}={\cal R}_{\mu\nu}^{\quad \;\mu\nu}$
and matching with  the Robertson-Walker result
\begin{equation}
{\cal R}=6\biggl(\frac {\ddot a}a + \Bigl( \frac {\dot a} a\Bigr)^2 + \frac k {a^2}\biggr)\;,
\label{k14}
\end{equation}
$\chi$ is fixed as
\begin{equation}
\chi = \frac{\dot a}{\sqrt{1-kr^2}}\;.
\label{k15}
\end{equation}

For simplicity we set all components of $\Theta^{\mu\nu}$ equal to zero except for
\begin{equation} \Theta^{tr}  = -\Theta^{rt}\equiv \theta \;.\label{k16}\end{equation}

Once again we are skipping the expressions of the NC extensions of vierbein \cite{stern} and only reproduce the  NC corrected FRW metric \cite{stern},
\begin{widetext}
\begin{eqnarray}
&&g_{tt}=-1+\frac{\theta^2}{16c^4} \left(6\ddot{a}^2+5\dot{a}\dddot{a} \right)+O(\theta^4) \nonumber\\
&&g_{rr}=\frac{a^2}{1-kr^2}-\frac{\theta^2}{16c^4} \Bigg(\dot{a}^4+13 a \ddot{a}\dot{a}^2 +12 a^2\dot{a}\dddot{a}+16 a^2 \ddot{a}^2 \Bigg)+O(\theta^4) \nonumber\\
&&g_{\theta\theta}=r^2 a^2+\frac{\theta^2}{16c^2} \Bigg[5\dot{a}^2+4 a \ddot{a}-\frac{a}{c^2} \Bigg(8 a \ddot{a}^2 +9\dot{a}^2\ddot{a}+4 a\dot{a}\dddot{a} \Bigg)r^2 \Bigg]+O(\theta^4) \nonumber \\
&&g_{\phi\phi}=\sin^2\theta g_{\theta\theta},
\end{eqnarray}
\end{widetext}
where  $c$ is velocity of light. Notice the well known fact \cite{ncgr} that all NC corrections start from $O(\theta^2)$ which is a characteristic feature of NC gravity.
We end this section with a cautionary note regarding the limitations of this NC extension framework, as mentioned by the authors of \cite{stern}. In particular the scale factor might receive additional NC corrections arising from NC matter couplings to Einstein gravity.

\section{Non-Commutative mini-superspace action}	
\label{sec-4}

It is plausible to argue that quantum  effects dominated in the early universe but without a fundamental theory of quantum gravity one has to contend with the Wheeler-de Witt form of quantum mechanics of the universe that works with a highly reduced system of finite number of degrees of freedom: the minisuperspace. It was shown in \cite{kief} that under certain criteria, fortunately satisfied by FRW metric, it is allowed to drastically reduce the degrees of freedom count at the action level thus  treating the universe itself as a mechanical system eventually leading to the Wheeler-de Witt version of Schr\"{o}dinger equation for the universe. However, we will stick to the classical scenario since  the NC generalization of FRW model, derived in  \cite{stern}, that we work with can have some quantization ambiguities  \cite{allen}. Importance of noncommutativity in High Energy Physics was emphasized by  Seiberg and Witten \cite{sw} who showed that in certain low energy limits open string dynamics, ending on $D$-branes in the presence of external two-form  gauge fields, was equivalent to a NC quantum field theory. Various aspects of NC space (time) effects on quantum systems as well effects on analogous non-canonical (Poisson) algebra on classical systems have been investigated (for reviews see  \cite{ncrev}).

In the present paper we show that a natural example of SSB in velocity space leading to an extended form of CTC is the FRW standard model of cosmology endowed with  NC correction \cite{stern}. The TC structure induces an interesting (but well studied) cosmological behavior such as bouncing or cyclic cosmology,  without any extra matter degrees of freedom.\\
{\it{NC-corrected minisuperspace action:}} The NC corrected FRW metric is given by \cite{stern}  where only $\Theta^{rt}=-\Theta^{tr}=\theta$ component of $\{x_\mu, x_\nu \}=\Theta_{\mu\nu }$ is non-zero and we keep $c$, velocity of light, explicit.  The generic form of the Einstein-Hilbert action $\mathcal{A} = \frac{c^4}{16 \pi G} \int (R - 2 \Lambda) \sqrt{-g}\; d^4x $ in the minisuperspace reduction reads\\
\begin{eqnarray}
\mathcal{A} = \sigma \int ~dt \Bigg [ \left( -a \dot{a}^2  + \kappa a - \frac{\Lambda}{3} a^3 \right) - \beta_1 \frac{\dot{a}^2}{a} + \beta_2 \frac{\dot{a}^4}{a} \nonumber\\-\beta_3 \frac{\dot{a}^6}{a} + \alpha _1 a \dot{a}^2 + \alpha_2 a \dot{a}^4 \Bigg] \label{action}
\end{eqnarray}
where NC corrections are introduced via $\theta$ in the numerical parameters,
\begin{eqnarray}
&&\sigma = \frac{c^2L^3}{2G} ,\;\;\;\; \beta_1 = \left( \frac{5 \theta^2}{16 L^4\rho }\right) , \;\;\;\; \beta _2 = \left( \frac{\theta^2}{4L^2 c^2}\right) ,\;\;\;\; \\
&&\beta_3 = \frac{\theta^2}{96 c^4},\;\;\;\; \alpha_1 =  \left(\frac{3 \Lambda \theta^2}{16L^2c^2}\right) ,\;\;\;\; \alpha_2 = \frac{317}{96 c^4} \Lambda \theta^2 \;\;\;\; .
\end{eqnarray}
In our model the dimensions of the physical quantities and parameters in mass ($m$), length ($l$) and time ($t$) units  are as follows:
 $$ [\mathcal{A}] =\frac{ml^2}{t},~ [G]=\frac{l^3}{mt^2},~ [a]=0,~ [\theta]=l^2,~ [\Lambda ]=[\kappa ]=\frac{1}{t^2}. $$
The parameter $[L]=l$  denotes the size of the comoving spatial coordinates and
$L^3$ arises from their integration when   the field theory is reduced to a quantum
mechanical system. Conventionally one sets $L=1$ by normalizing the scale factor $a$. However the situation is more subtle here since the NC correction terms generate  inhomogeneity and does not allow $L^3$ to be uniformly factored out. This forces us to keep $L$ explicit for the time being.

Some comments about the action in (\ref{action}) are in order. We have considered a closed ($\kappa =1 $)  universe in the canonical sector of the action but have not considered $\kappa$-dependent  terms in the NC $O(\theta^2)$ contributions, primarily for convenience. (We also believe that $\kappa$-dependent NC corrections in the metric \cite{stern} will not change our major conclusions significantly.)  Secondly, we have followed an approximation scheme suggested in \cite{stern} that drops $\ddot a(t)$ and higher time derivatives, again from  NC correction terms only. Finally,  surface terms are dropped making the action a functional of $a,~\dot a $ to allow a Lagrangian description. We note that, $\rho <<1 $ is a numerical parameter (to be fixed later), that appears only in $\beta_1$ as a regularization that introduces a NC-induced fuzziness ($\sim$ Planck length) in the minisuperspace volume. It is immediately clear that the action (\ref{action}) possesses TC behavior but with novel features compared to the models in \cite{wil,wil11} primarily due to the $\dot a^6$   term and  overall for more complicated $a$ dependence.

 \begin{figure}
 \begin{center}
 \includegraphics[width=0.45\textwidth]{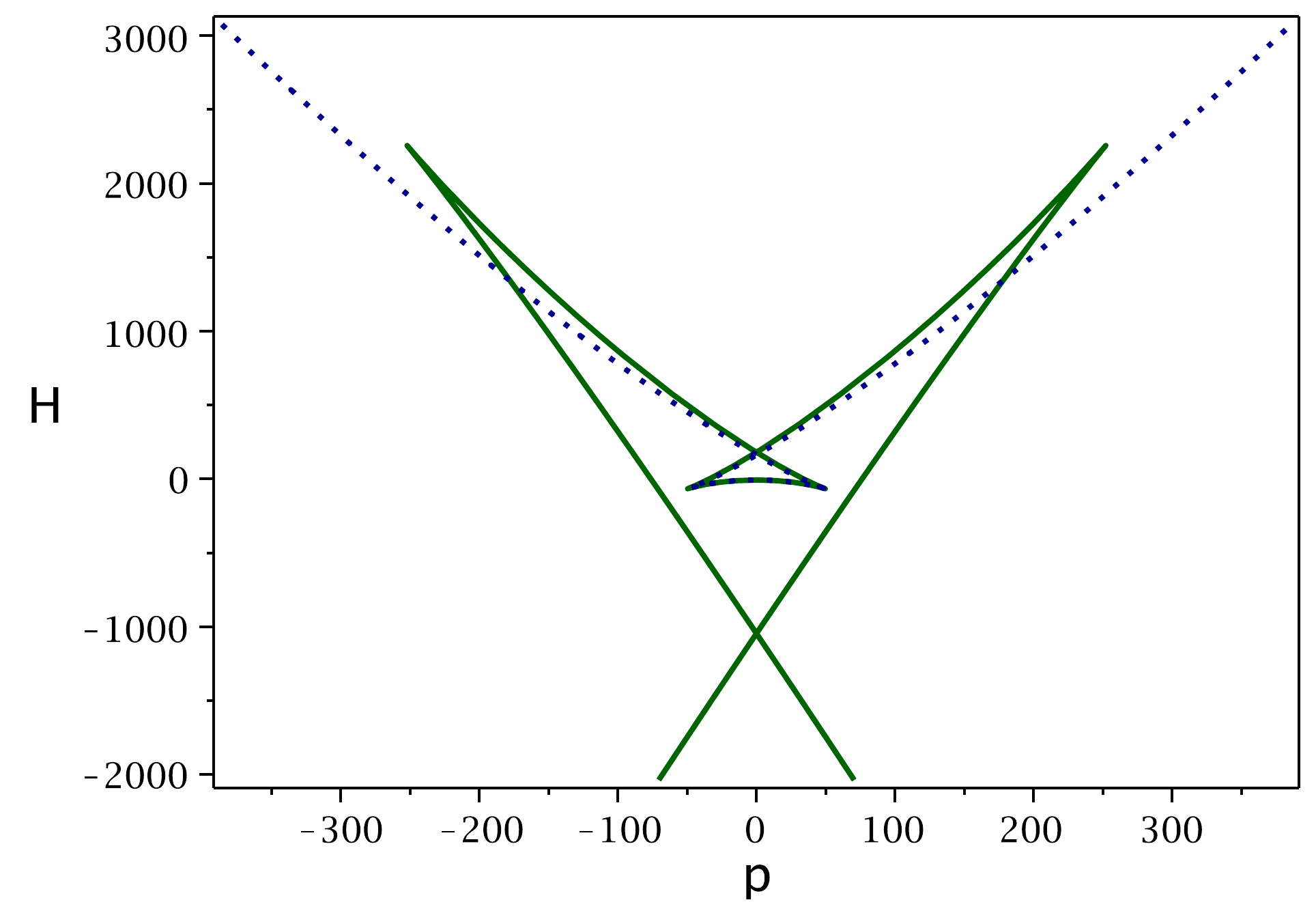}
 \caption{The figure displays the $p$ vs. $H$ diagram where solid and dotted lines show \textit{batwing} and \textit{swallowtail} features respectively. }
 \label{fig-p-vs-H}
 \end{center}
 \end{figure}

Let us identify $L=c/\Lambda $ (to be justified presently) and rewrite (\ref{action}) in a compact form,
\begin{eqnarray}
\mathcal{A} = \sigma \int ~dt \left(  -A \dot{a}^2  + B\dot{a}^4 -C\dot{a}^6 + \kappa a-\frac{\Lambda}{3}a^3  \right), \label{action1}
\end{eqnarray}
where $A=a+\frac{\nu}{16} \left(\frac{5}{\rho a}-3a \right)$; $B=\frac{\nu}{4\Lambda}\left(\frac{1}{a}+\frac{317 a}{24} \right)$; $C=\frac{\nu}{96\Lambda^2 a}.$
We have introduced a dimensionless universal parameter $\nu=\frac{\theta^2\Lambda^2}{c^4}$.
 Defining the canonical momentum 
 \begin{equation}
  p=(\partial \mathcal{L})/(\partial \dot a)=\sigma (-2A\dot a+4B\dot {a}^3-6 C\dot {a}^5),
\label{p}
 \end{equation}
 our cherished Hamiltonian yields,
  \begin{equation}
 H=p\dot a - \mathcal{L}  = \sigma  \left(  -A \dot{a}^2  + 3B\dot{a}^4 -5C\dot{a}^6 - \kappa a+\frac{\Lambda}{3}a^3  \right).
 \label{h}
 \end{equation}
 This Hamiltonian constitutes our primary result which we now study from the CTC perspective. First of all note that $a$ being the scale factor is positive. This makes $B,C$ positive and $A$ will be positive provided  the small parameter $\rho $ is less that  $5\Lambda /3$.
 
 \section{Cosmological time crystal} 
 \label{sec-5}
 
 Let us now study the novelties of our proposed TC model, where for the time being, we treat $\nu$ and $\Lambda$ simply as arbitrary parameters. Later we will deal  with  the cosmological scenarios. 
 
 {\it{Cusp structure}}: A   richer cusp structure is revealed from the $p$ vs. $ H$ diagram in Fig. \ref{fig-p-vs-H}. Following \cite{wil11} the plot of $p$ vs. $H $ shows that because of the $\dot {a}^6$-term, $\partial p/\partial \dot a =0$, is a fourth order equation in $\dot a$ having four solutions as  seen from Fig. \ref{fig-p-vs-H}: the dotted blue line resembles the profile of \cite{wil11} where  $\dot a^6$ contribution is negligible whereas the solid deep-green line shows the new structure with   significant $\dot {a}^6$ contribution.   We name this new structure as {\it{batwing}} catastrophe that reduces to the swallowtail catastrophe (obtained in \cite{wil11}) if the contribution of the  $\dot{a}^6$-term becomes negligible. On the other hand the cusp structure completely washes out if the $\dot {a}^6$-term dominates even slightly, leaving an inverted symmetric V-shaped diagram with the vertex at $H=0$.

 \begin{figure*}
 	\includegraphics[width=0.42\textwidth]{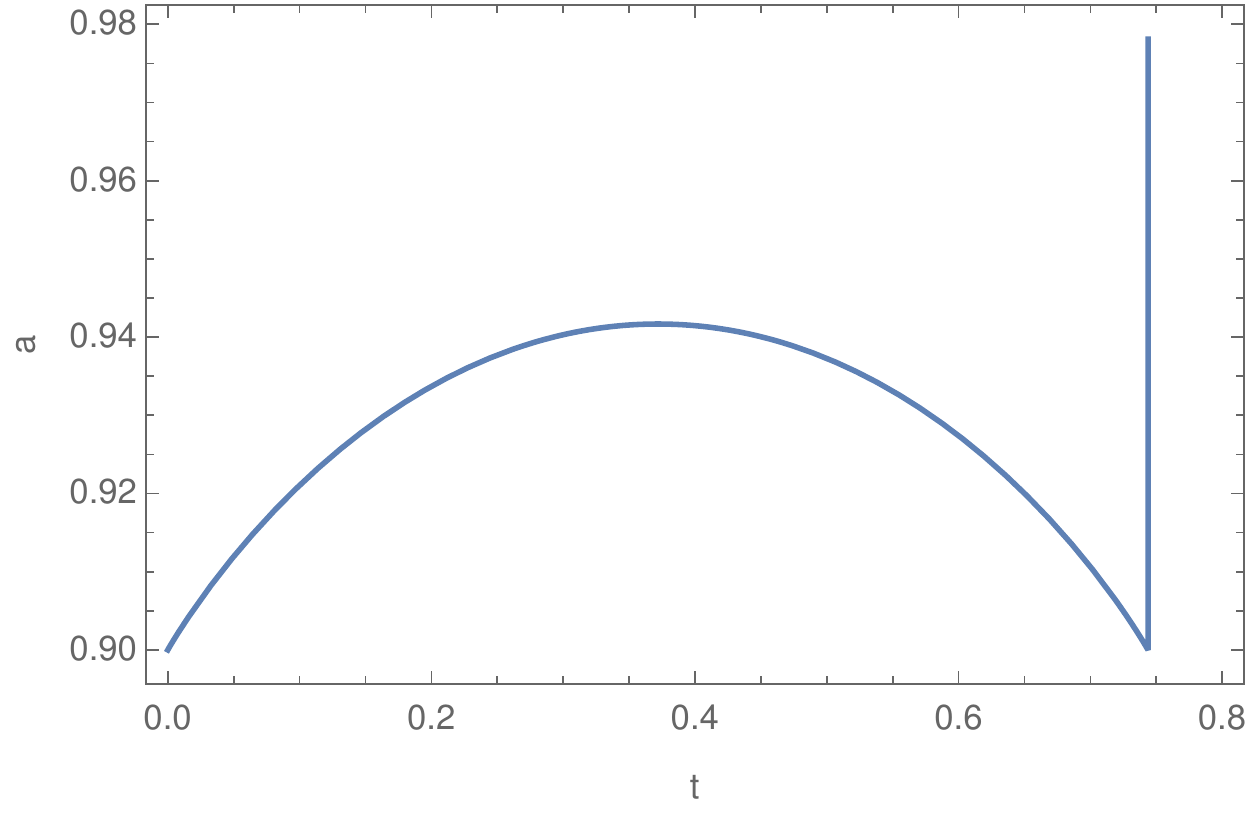}
 	\includegraphics[width=0.40\textwidth]{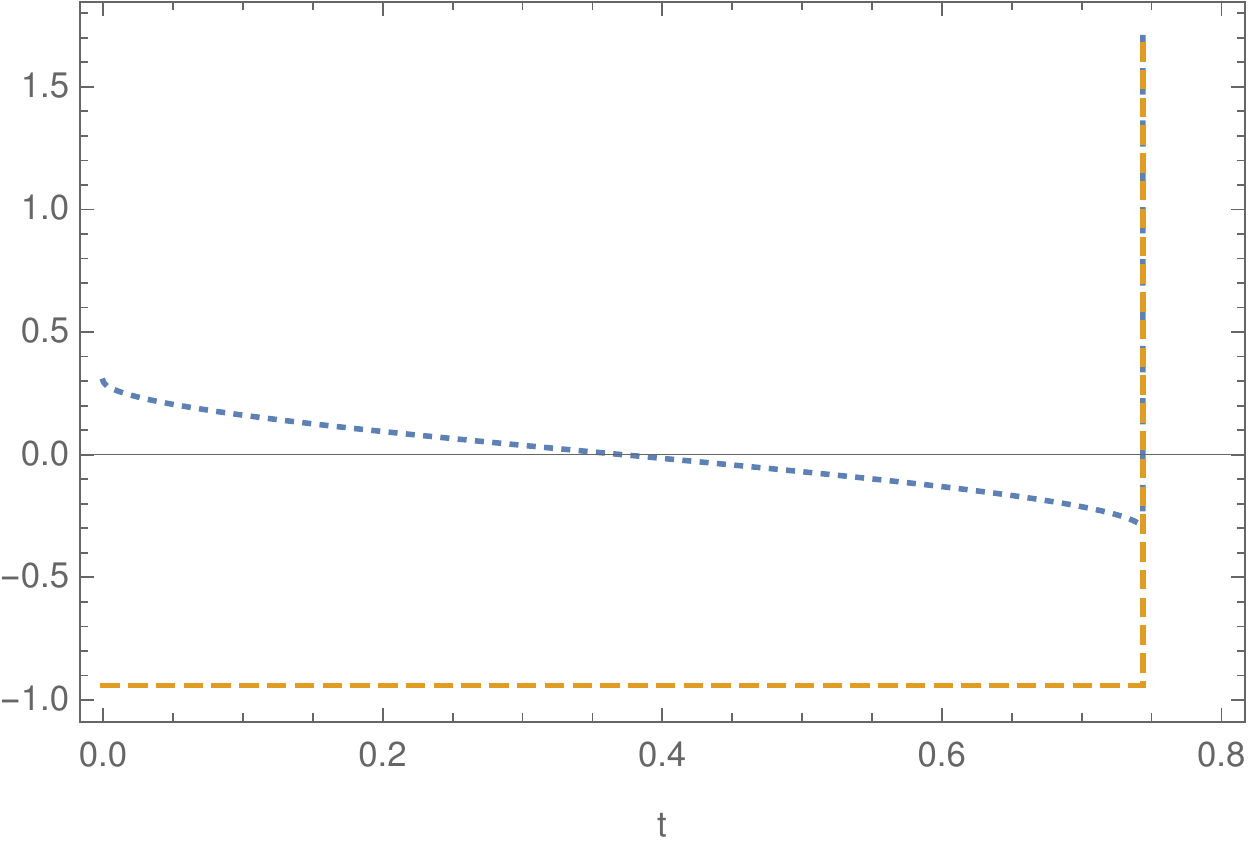}
 	\caption{The dynamics of the eqn. \ref{leq}  for some particular choices of the model parameters has been presented. Left panel shows the evolution of the scale factor while the right panel shows the expansion rate $\dot{a}$ (blue dotted curve) and the total energy $H$ (brown dashdot curve).}
 	\label{fig-a-dot-four-singul}
 \end{figure*}

 %\begin{figure}
 %\includegraphics[width=0.32\textwidth]{Scale_Factor_a-dot-four.pdf}
 %\includegraphics[width=0.30\textwidth]{Energy_and_expansion-rate_a-dot-four.pdf}
 %\caption{The dynamics of the equation (\ref{leq}) for some particular choices of the parameters. The upper panel shows the evolution of the scale factor while the lower panel shows the expansion rate $\dot{a}$ (blue dotted) and the total energy $H$ (brown dashdot). }
 %\label{fig-a-dot-four-singul}
 %\end{figure}

 \begin{figure}
 \includegraphics[width=0.42\textwidth]{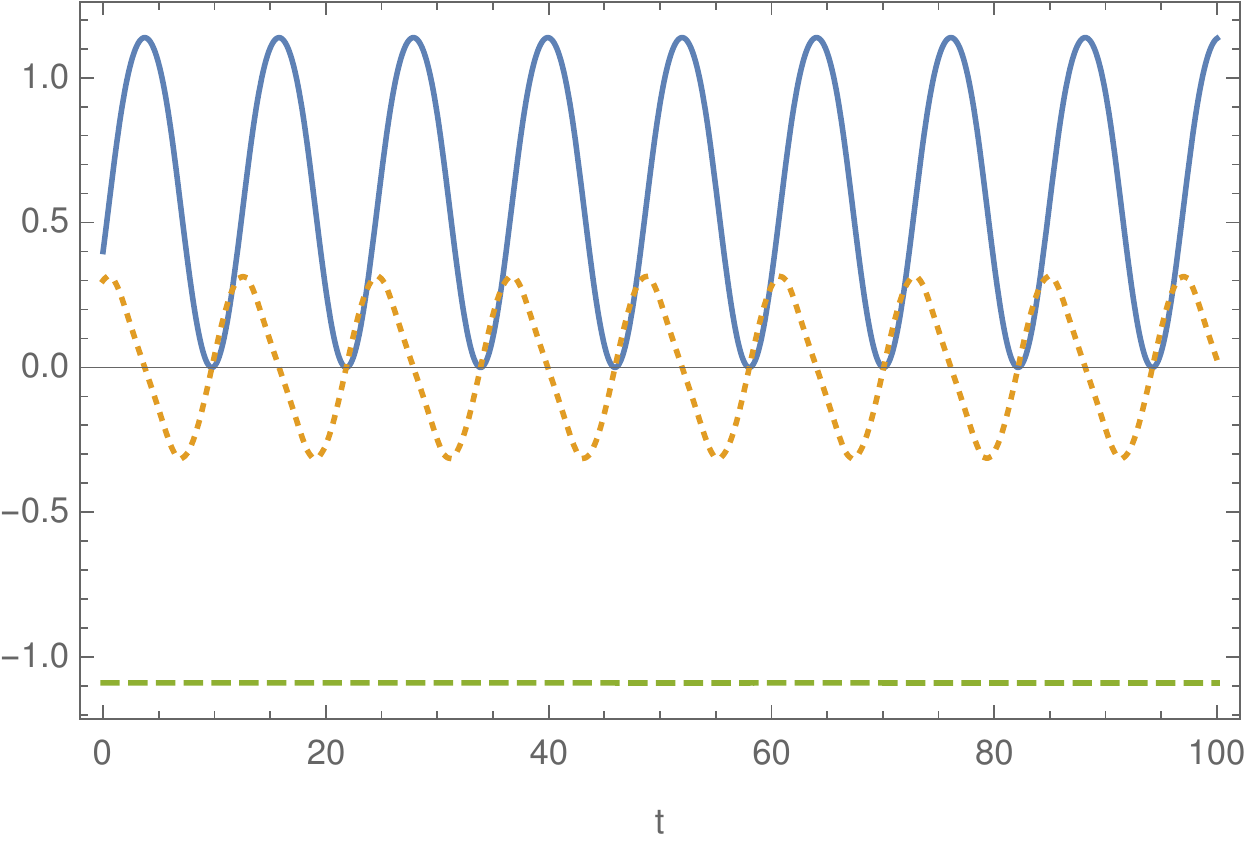}
 \caption{ The profile of the scale factor $a$ (solid curve), its time derivative $\dot{a}$ (dotted curve), and the total energy $H$ (dashdot curve) for the dynamical system in presence of the $\dot{a}^4$-term only  (\ref{leq}) have been displayed for some specific choices of the parameters. }
 \label{fig-a-dot-four-no-singul}
 \end{figure}

  \begin{figure}
 \includegraphics[width=0.42\textwidth]{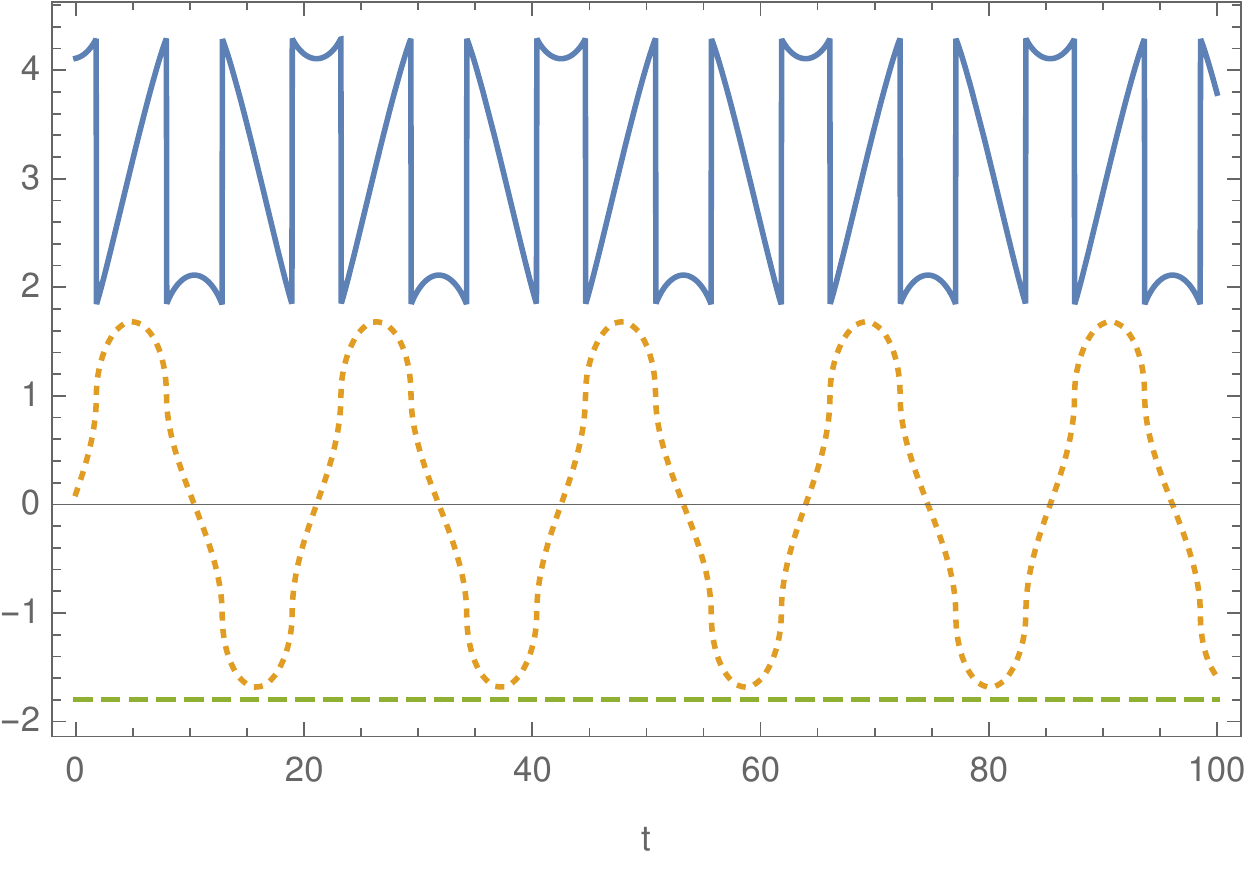}
 \caption{The profile of the scale factor $a$ (solid curve), $x$ (dotted curve) and $H$ (dashdot curve) have been shown for the system of equations (\ref{eqm0.1}) and (\ref{eqm0.2}) for some specific choices of the parameters.}
 \label{fig-a-dot-four-with-mu}
 \end{figure}
 
 \begin{figure}
 \includegraphics[width=0.42\textwidth]{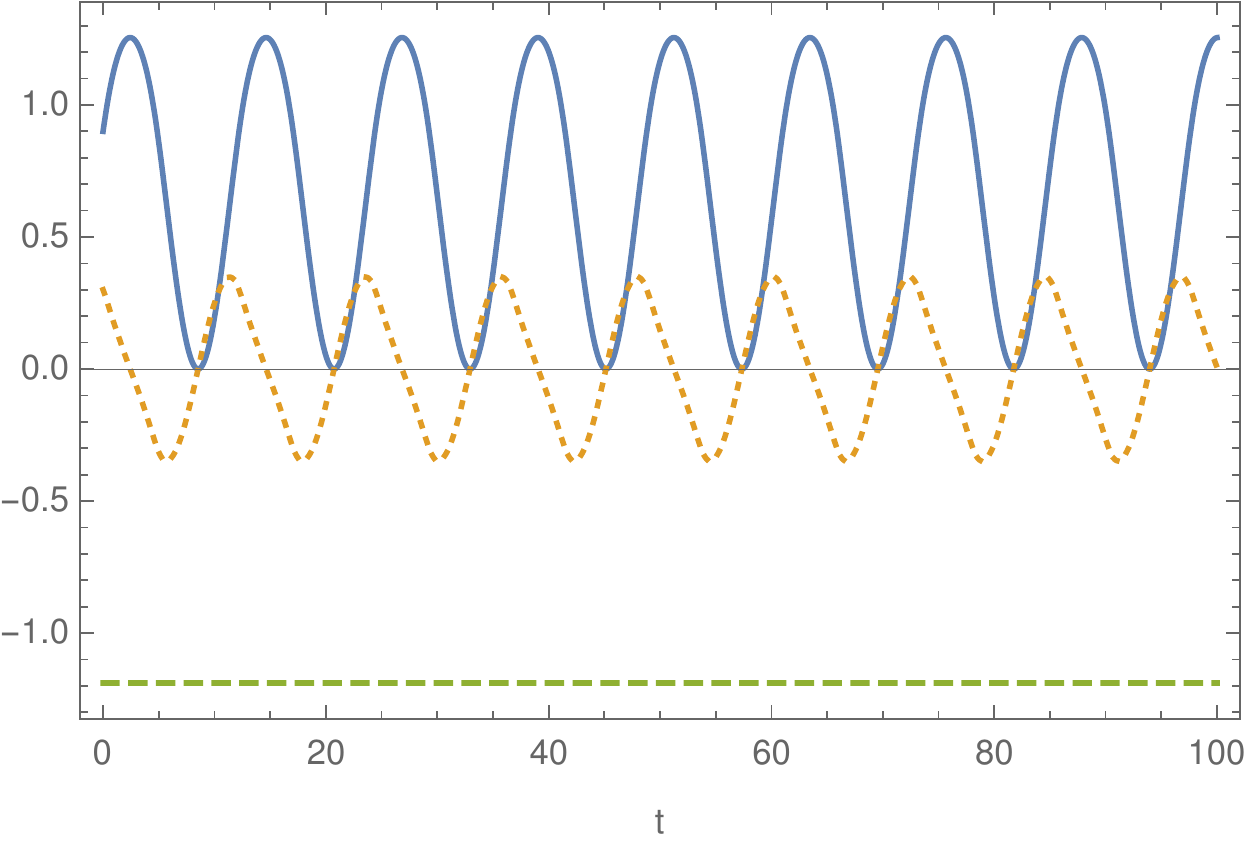}
 \caption{The profile of the scale factor $a$ (solid), $x$ (dotted) and $H$ (dashdot) for the equation (\ref{eqm}) have been presented for some specific choices of the parameters.}
 \label{fig-a-dot-six-no-mu}
 \end{figure}
 
 \begin{figure*}
 \includegraphics[width=0.42\textwidth]{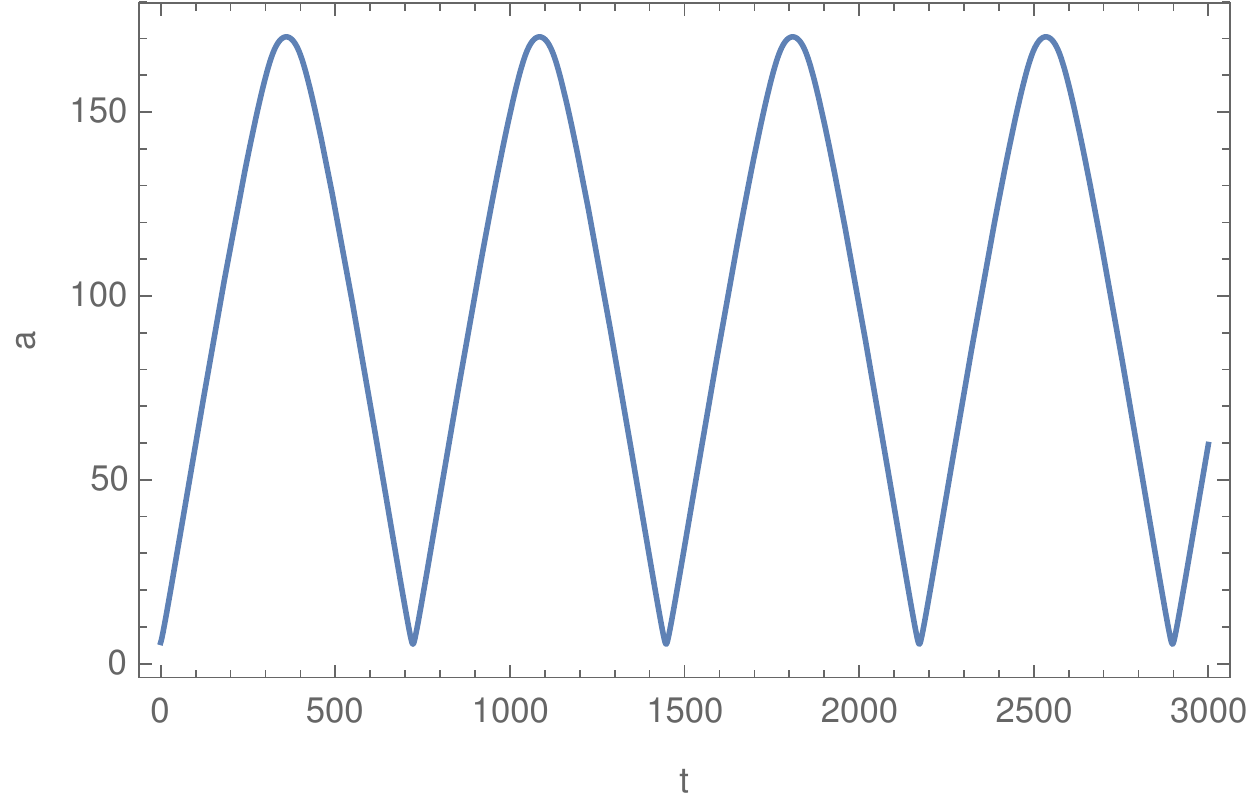}
 \includegraphics[width=0.39\textwidth]{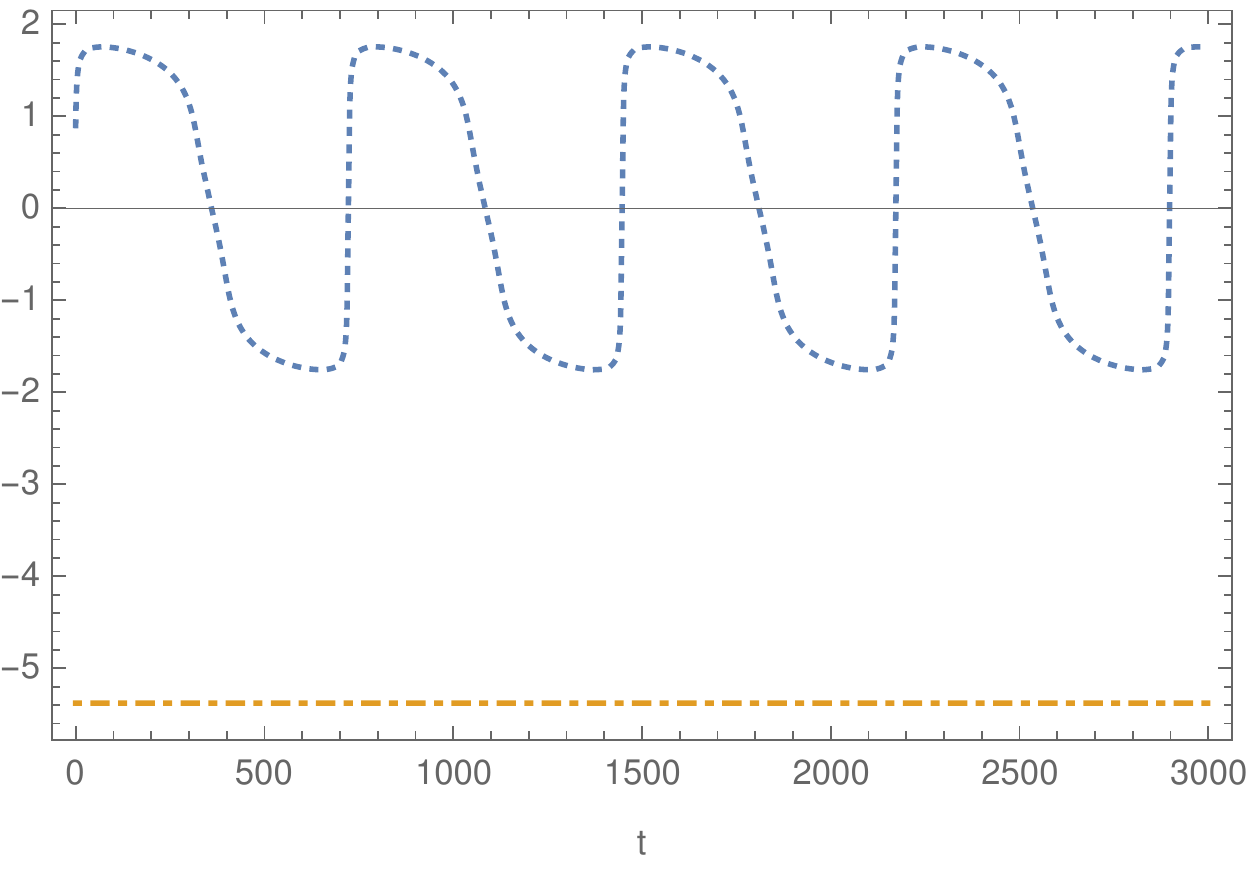}
 \caption{ Dynamics of the regularized CTC in presence of $\dot{a}^6$ term. Left Panel: Profile for the scale factor $a$ that represents a sequence of expansion of contraction of the universe. Right Panel: Profile for $x$ (dotted; Sisyphus behavior) and the total energy of the system $H$ (dashdot).}
 \label{fig-a-dot-six-with-mu}
 \end{figure*}

 {\it{Effective planar model without  $\dot a^6$-term}}: Let us start by dropping the $\dot a^6$-term (which we will justify later in the cosmological context). A naive plot of  $a(t)$ and $\dot a(t)$ from the Lagrangian equation of motion, with $A'=\partial A/\partial a, ..$, 
 	\begin{equation}
 	2\ddot a(\dot a ^2-A)+\dot a ^2 (3B'\dot a ^2-A')-1+\Lambda a^2=0,
 	\label{leq}
 	\end{equation}
 	is given in Fig. \ref{fig-a-dot-four-singul} where $a(t)$, $\dot{a}(t)$, $H$ are plotted against the cosmic time $t$.  From this profile we find a single hump before it hits a singularity{\footnote{In Fig. \ref{fig-a-dot-four-singul} the singular nature of $a(t)$ is shown where in the left panel $a(t)$ hits a singularity around $t\sim 0.8$. This can be interpreted as a big rip type of singularity (a future type of singularity often appearing in the cosmological models) \cite{Nojiri:2005sr} in the cosmological context. }}. However, quite interestingly, we have discovered a particular set of parameter values and initial conditions that give rise to a stable oscillating $a$ that hits a singularity after a very long time (see Fig. \ref{fig-a-dot-four-no-singul}).  Thus, one may infer that the cosmological model framed by TC may allow a cyclic nature of the universe. While we believe that further investigations toward this direction is necessary, possibly to understand the nature of entropy transfer from one cycle to another since during the evolution of the universe. This effectively may reveal whether the model is geodesically past incomplete or not.

{\it{Regularized CTC}}: In the generic case  we follow the elegant route of \cite{wil11} that shows a way of removing the singularity by introducing a regulator in the form of $\mu $ in an alternative Lagrangian,
\begin{equation}
\mathcal{L}=\frac{\mu}{2}\dot x^2+f(x)J(a)\dot a-g(x)(1+K(a))-V(a),
\label{L1}
\end{equation}
with
\begin{equation}
f(x)=\frac{x^3}{3}-x,~g(x)=\frac{x^4}{4}-\frac{x^2}{2}
\label{0a}
\end{equation}
 as prescribed in \cite{wil11} and in addition, the quantities $K$, $J$ in (\ref{L1}) are,
\begin{equation}
1+K(a)=\frac{A^2}{3B},~J(a)=\sqrt{\frac{2A^3}{3B}}.
\label{0b}
\end{equation}
The Lagrangian  (\ref{L1}) mimics a planar charged particle in an external potential. The equations of motion from (\ref{L1}) are
\begin{eqnarray}
\mu \ddot x-f'(x)J(a)\dot a+(K(a)+1)g'(x)=0,\label{eqm0.1}\\
J(a)f'(x)\dot x+g(x)K'(a)+V' (a) =0.
\label{eqm0.2}
\end{eqnarray}
Similar to \cite{wil11}, in the present case for $\mu =0$, the coupled set above reduces to (\ref{leq}).  
Fig. \ref{fig-a-dot-four-with-mu} shows the behavior of the scale factor $a$ (blue solid line), $x$ (brown dotted line) and the total energy of the system $H$ (green dashed line) with the evolution of the cosmic time, $t$. Although the present model is similar to that of \cite{wil11} from the TC point of view (both having only $\dot a^2$ and $\dot a^4$ terms),  however, the $a$-dependence is much more complicated in our case.  This is reflected  in our $x$ profile (Fig. \ref{fig-a-dot-four-with-mu}) that contains both positive and negative values (whereas \cite{wil11} has either positive or negative values) with the sharp edges separated by smooth curves. Compared to \cite{wil11} our model describes a {\it{doubled}} Sisyphus dynamics.

 {\it{Effective planar model with  $\dot a^6$-term}}: We now present our most important result where the full TC model is studied. Generic solutions of the equation of motion,
 	\begin{eqnarray}
 	2\ddot a(A-6B{\dot a}^2+15 C{\dot a}^4) +{\dot a}^2(A'-3B'{\dot a}^2+5C'{\dot a}^4) \nonumber\\+1-\Lambda a^2=0,
\label{eqm}
\end{eqnarray}
with $A'=\frac{dA}{da}$, $B'=\frac{dB}{da}$, and $C'=\frac{dC}{da}$,  have singularities as expected.	
 A very surprising and interesting result is that we have discovered a small parameter window in which the scale factor $a(t)$ oscillates smoothly without any singularity (as far in time as we have checked) and with constant energy indicating a cyclic universe. This is shown in Fig. \ref{fig-a-dot-six-no-mu}.  However, we stress that further investigations are necessary because the initial conditions on some parameters, as we observed, play an essential role in determining the nature of the scale factor. 
 
 In order to construct the $\mu$-regularized model we try  with a simple generalization. Keeping the same form of $\mathcal{L}$ as in (\ref{L1})
 $$
 \mathcal{L}=\frac{\mu}{2}\dot x^2+f(x)J(a)\dot a-g(x)(1+K(a))-V(a), 	
 $$ but with new forms of $f(x),~g(x)$ as,
 	\begin{equation}
 	g(x)=\frac{x^6}{6}-\frac{x^4}{4}+\frac{x^2}{2},~~f(x)=\frac{x^5}{5}-\frac{x^3}{3}+x
 	\label{6b}
 	\end{equation}
 	we find the same set of equations of motion as in (\ref{eqm})  with the identifications
 	\begin{equation}
 	  	 1+K(a)=\frac{A^2}{3B}, ~~
 	 J(a)=\sqrt{\frac{2A^3}{3B}},~~30 C=\frac{J^6}{(1+K)^5}.
 	\label{0c}
 	\end{equation}
The last identification shows that the regularized form for $\dot a^6$-model is not entirely satisfactory as it fails to generate the independent form of $C$ given in (\ref{eqm}). This suggests that a more elaborate version of the $\mu$-regularized model is needed to faithfully represent the parent $\dot a^6$-model. This has not been pursued here. However, the model with (\ref{6b}) is indeed an example of a TC with $\dot a^6$ whose equations of motion are 
	\begin{eqnarray}
	\mu \ddot x -f'(x)J(a)\dot a +g'(x)(1+K(a))=0,\nonumber\\
	f'(x)J(a)\dot x+g(x)K'(a)+V'(a)=0,	
	\label{6a}
	\end{eqnarray}
	with solutions  displayed in Fig. \ref{fig-a-dot-six-with-mu} with Sisyphus-like behavior.

\section{Cosmological time crystal and a small Cosmological constant $\Lambda_{\rm eff}$}
\label{sec-7}
 
\begin{figure}
 \includegraphics[width=0.3\textwidth]{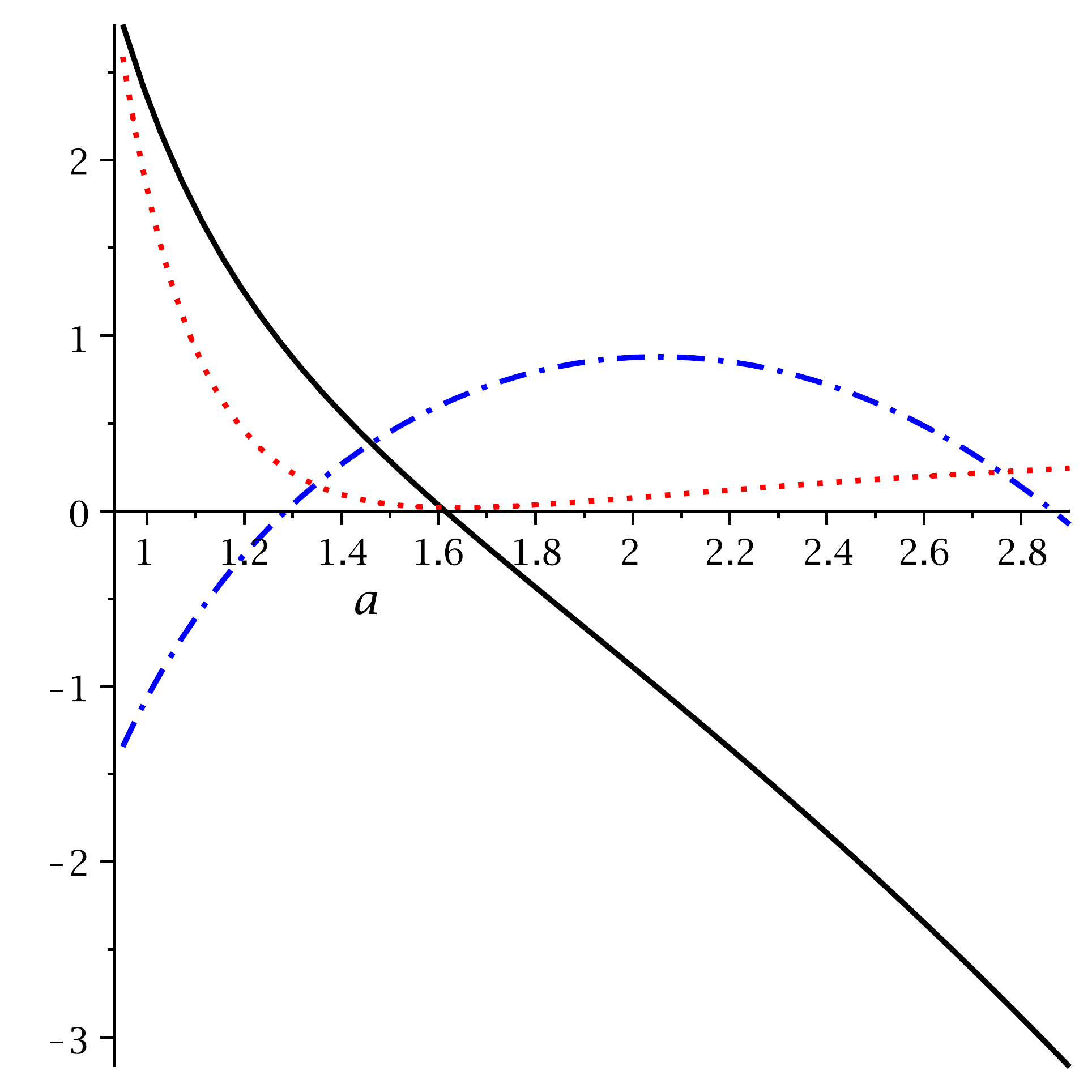}
 \caption{ Qulitative evolutions for the variation of the effective potential,i.e., $\partial V_{\texttt{eff}}/\partial a$ (solid line); the effective cosmological constant $\Lambda_{\texttt{eff}}$ (dotted line) and the Hamiltonian having its ground state energy $H_{\rm condensate}$ (dashed line) have been shown with the evolution of the universe in terms of the scale factor $a$. }
 \label{Fig-Lambda}
 \end{figure}

	Let us derive the condensate energy where we need the condensate values for $a_0$ and $\dot a_0$ that minimize the Hamiltonian. The Hamiltonian $H$ for the action (\ref{action1}) without the $\dot a^6$ term, can be written in the form			
	\begin{eqnarray}
	&&H=\sigma \left[3B \left(\dot a ^2-\frac{A}{6B} \right)^2+V_{{\rm eff}} \right],\\	
	&&V_{{\rm eff}} = \left( a-\frac{\Lambda}{3}a^3-\frac{A^2}{12B} \right).\label{4}
	\end{eqnarray}
	This approximation is justified \cite{stern} as $\dot a^6$-term is induced by higher order derivatives. Since $B$ is always positive, $H$ will be minimized for 
	\begin{equation}
		\dot {a}_0 =\pm \sqrt{\frac{A(a_0)}{6B(a_0)}}=\pm \sqrt{\frac{A_0}{6B_0}}
	\label{0p}
	\end{equation}
 with $a_0$ obtained from $\frac{\partial V_{\rm eff}}{\partial a}\mid_{a_0}=0$. The ground state energy will be 
 \begin{equation}
 H_{condensate}=- \frac{A_0^2}{12 B_0}+a_0-\frac{\Lambda}{3}a_0^3, $$$$ A_0=A(a_0), B_0=B(a_0).
 \label{0u}
 \end{equation}
    In Fig. \ref{Fig-Lambda}  the solid line ($\partial V_{\texttt{eff}}/\partial a$) cuts the $a$-axis at $a_0$, which incidentally is greater than $a=1$ (i.e., in future). We define
   \begin{equation}
    \Lambda_{\rm eff}= \Lambda - \frac{3}{a_0^2}+ \frac{A_0^2}{4 B_0 a_0^3}
   \label{0y}
   \end{equation}
    and  the value of $\Lambda (a_0)=\Lambda_{\rm eff}$ can be arbitrarily small (as seen from Fig. \ref{Fig-Lambda}) even with a large value of $\Lambda$: the value of Cosmological Constant  gets renormalized in a sense. Furthermore, the interpretation of $H$ as  energy of the condensate is consistent within the range of $a$ where $H_0$ is positive.

 \section{Noncommutative Friedmann equation (without $\dot a ^6$ term)}
 \label{sec-6}
 
 Friedmann equations are derived from a generic minisuperspace model in a straightforward way: one equation is the Lagrangian equation of motion for $a(t)$ and the other equation comes from simply putting the Hamiltonian to zero since the Hamiltonian being one of the generators of diffeomorphism symmetry,  comes as a factor of the lapse function. However, in the present case, due to two forms of  SSB, (in velocity $\dot a(t)$ and coordinate $ a(t)$ sectors), one has to shift both $a(t),~\dot a(t)$ by their respective condensate values to get the  action in terms of new variables where the condensate does not appear any more with the degrees of freedom vanishing in the ground state.  This amounts to using the variable $b(t)$ defined by  the transformations 
 \begin{equation}
 b(t) = a(t)-a_0,~~\dot b(t) = \dot a(t)-\dot a_0
 \end{equation}
 in the action. The new action yields the following set of equations,		 
 \begin{widetext}
 	\begin{eqnarray}
 	&&\frac{\dot{b}^2}{b^2} + \frac{\bar{k}}{b^2} = \frac{\Lambda (b+a_0)^3}{3 b^2 f} + \frac{1}{b^2} \, \frac{g}{f},\label{MFE1}\\
 	&&\ddot{b} = \frac{1}{h} \Bigg[ \frac{b^2 f}{b+a_0} \left( \frac{g}{b^2 f} + \frac{\Lambda (b+a_0)^3}{3 b^2 f} -\frac{\dot{b}^2}{b^2}\right) - \Lambda (b+a_0)^2 + \left(\dot{b}^2 - \dot{a}_0^2 \right) \left(1 -\frac{3\nu}{16} - \frac{5\nu}{16\rho (b+a_0)^2}\right) \nonumber\\ &+& \frac{\nu}{\Lambda} \left( \dot{b} + \dot{a}_0\right)^3 \left( \frac{\dot{b} + \dot{a}_0}{4} -\dot{b}\right) \left( \frac{317}{24} - \frac{1}{(b+a_0)^2}\right) \Bigg],\label{MFE2}
 	\end{eqnarray}
 \end{widetext}
 where $\bar{k} = k (b+a_0)/f$ is the effective  curvature scalar and the functions $f$, $g$ respectively have the forms,
 \begin{widetext}
 	
 	\begin{eqnarray}
 	&&f = (b+a_0) \left( 1-\frac{3\nu}{16} - \frac{317 \nu \dot{a}^2}{16 \Lambda} \right) + \frac{\nu}{b+a_0} \left( \frac{5}{16\rho} - \frac{6 \dot{a}^2}{4 \Lambda} \right),\nonumber\\
 	&&g = \dot{a}^2 \left[ \left(1- \frac{3\nu}{16}\right) (b+a_0) + \frac{5\nu}{16 \rho (b+a_0)}\right] + \frac{\nu}{4 \Lambda} \left( 3 \dot{b}^4 + 8 \dot{b}^3 \dot{a}_0 - \dot{a}_0^4\right) \left( \frac{1}{b+a_0} + \frac{317}{24} (b+a_0)\right),\nonumber\\
 	&&h = \frac{3\nu}{\Lambda} \left(\dot{b} + \dot{a}_0\right)^2 \left(\frac{1}{b+a_0} + \frac{317}{24} (b+a_0)\right)- \left[ \left(1-\frac{3\nu}{16}\right) \left(b+a_0\right)+ \frac{5\nu}{16 (b+a_0)}\right].
 	\end{eqnarray} 
 \end{widetext} 	
 Here,   (\ref{MFE1}) is identified as the Hubble equation and  (\ref{MFE2}) is the acceleration equation. One may note that the effective curvature scalar $\bar{k}$  becomes variable and moreover, from the right hand side of (\ref{MFE1}),  an effective cosmological constant which now gains a time dependent character is also realized. In other words, the noncommutative corrections in the FRW model with a time independent cosmological constant can produce an effective scenario where the cosmological 
 	constant runs with the evolution of the universe. 
 
\section{Discussion} 
\label{sec-8}

Let us summarize our results. We have considered an existing model of generalized FRW metric endowed with noncommutativity corrections. This extended FRW model gives rise to a new form of time crystal behavior with a specific form of singular behavior (\textit{batwing-catastrophe} $-$ the name coined by us) that can be compared with the simpler form of singularity $-$ swallowtail catastrophe, already encountered in  \cite{wil,wil11}. In addition to that, our model has a  doubly Sisyphus dynamics, as a dynamical model with generic parameter values, that can again be contrasted with \cite{wil11}. These qualitative differences arise essentially because our model has a more involved kinetic sector with the presence of exotic terms involving quartic and sextet order velocity contributions and field dependent factors multiplying the velocity terms (whereas previously studied models \cite{wil,wil11} had only quartic terms involving velocity and constant factors multiplying the velocity terms). 
	 
Since our time crystal model is a form of generalized FRW model, it is natural to term it as a Cosmological Time Crystal and study the consequences in cosmology. In minisuperspace framework adapted here, the system reduces to a mechanical one with the scale factor $a(t)$ emerging as the single dynamical variable.  Since $a(t)$ is associated with motion  periodic in time - Sisyphus dynamics - the model can be interpreted as a form of Cyclic Cosmology in a natural way. Apart from revealing a  new form of (Cosmological) Time Crystal, this is one of our  major findings. This   cyclic  behavior is depicted  in the eternally bouncing solutions depicted in Fig. \ref{fig-a-dot-four-no-singul} (for $\dot a^4$-model) and  Fig. \ref{fig-a-dot-six-no-mu} (for $\dot a^6$-model) that substantiate our view  that noncommutative gravity can represent cyclic cosmologies.  However, it should be stressed that our model of cyclic cosmology is qualitatively distinct from conventional cyclic cosmological models \cite{cycl, Novello, Lehners} since the latter involved matter fields put in from outside whereas our model is purely geometrical in the sense that no (matter) field is introduced from outside. The other difference in perspective is that in conventional cyclic cosmological approach, it is generally believed that Quantum Gravity contributions are not important. On the other hand in our scheme, even though the analysis of the generalized FRW model is entirely classical, the model itself can be thought of as an effective theory that contains essential noncommutative contributions originating  from Quantum Gravity effects. So far our analysis suggests that the noncommutative gravity can provide an interesting and physically motivated toy model. More work is needed before it can be elevated to a viable Quantum Gravity model.
	 
	 The other major success of our toy model approach is that it allows us to generate  an arbitrarily small positive cosmological constant in a natural way.  We have also derived a modified form of Friedman equation that has noncommutative contributions.
	 
	 There are many open problems that can be studied in the Cosmological Time Crystal framework.\\
	 (i)  Detailed analysis of the noncommutativity modified Friedman equation needs to be studied. \\
	 (ii)  Inclusion of matter degrees of freedom is another area to be looked at since one can conjure up more than one Time Crystal structures of  with Time Crystal behavior coming from the matter sector as well as the FRW (scale factor) sector. Also presence of matter will bring these systems closer to realistic cosmological models.\\
	 (iii)  Another very interesting avenue to explore is to analyze other forms of Time Crystal that can appear naturally in cosmology, in particular in $f(R)$ gravity (that will contain quartic and still powers of velocity, conducive for Time Crystal of the form studied in the present work) and in higher derivative gravity (where Time Crystals can form following the approach as desribed in \cite{sg}). We plan to pursue these areas in near future.

\section*{Acknowledgments} It is a pleasure to thank Anirban Saha, Pradip Mukherjee, Allen Stern and Krzysztof Sacha for helpful correspondence. We  thank Mark Hertzberg  for informing us about their works. We are grateful to the anonymous referee for constructive comments and also for suggesting us to look for other forms of Cosmological Time Crystal in $f(R)$ gravity. The work of P.D. is supported by INSPIRE, DST, India.

\end{document}